\documentclass[epj]{svjour}
\usepackage{amssymb}
\usepackage{amsmath}
\usepackage{graphicx}
\usepackage{textcomp}
\usepackage{amsfonts}
\usepackage[colorlinks=false, linkcolor=black]{hyperref}

\DeclareGraphicsExtensions{.pdf}
\ifx\pdftexversion\undefined\DeclareGraphicsExtensions{.eps}\fi

\begin{document}

\title{Laser diffraction by periodic dynamic patterns in anisotropic fluids}
\author{Thomas~John\inst{1,2}, Ulrich~Behn\inst{2}, and Ralf~Stannarius\inst{1,3}}
\date{Received: date / Revised version: date}
\institute{Institut f\"{u}r Experimentelle Physik I, Universit\"{a}t Leipzig,
Linn\'estrasse 5, \and Institut f\"{u}r Theoretische Physik, Universit\"{a}t
Leipzig, Vor dem Hospitaltore 1, D-04103 Leipzig, Germany \and Institut f%
\"{u}r Experimentelle Physik, Universit\"{a}t Magdeburg, Universit\"{a}%
tsplatz 2, D-39106 Magdeburg}

\abstract{ This paper describes the application of a laser diffraction
technique to the study of electroconvection in nematic liquid crystal cells. It
allows a real-time quantitative access to pattern wave lengths and amplitudes.
The diffraction profile of the spatial periodic pattern is calculated and
compared quantitatively to experimental intensity profiles. For small director
tilt amplitudes $\varphi$, the phase grating generated in normally incident
undeflected light and the first order term correction from light deflection is
derived analytically. It yields an $I\propto\varphi^4$ dependence of the
diffracted intensity $I$ on the amplitude of director deflections. For larger
director tilt amplitudes, phase and amplitude modulations of deflection of
light in the inhomogeneous director field are calculated numerically. We apply
the calculations to the determination of the director deflection and measure
growth and decay rates of the dissipative patterns under periodic excitation.
Real time analysis of pattern amplitudes under stochastic excitation is
demonstrated. }


\authorrunning{Th.~John et al.} \titlerunning{Laser diffraction by EHC  }

\maketitle

\section{Introduction}

Electrohydrodynamic convection (EHC) in nematic liquid crystals is one of
the standard systems of dissipative pattern formation. It has been studied
extensively during past decades. As a consequence of anisotropic properties
of nematic phase, the system is particularly rich in morphology. Among the
advantages of this system for experimental characterization are the easy
control of electric excitation fields, convenient time scales and the
straightforward observation techniques.

The equations describing the fundamental mechanism yield two dynamic
regimes: conduction and dielectric structures. In addition to the primary
instability toward simple roll patterns with wave vectors normal or inclined
to the preferential alignment of the director (optic axis) of the system, a
variety of secondary instabilities have been described. Besides the
investigation of arrays of parallel rolls, scientific interest recently
focussed on defect structures and localized convection states \cite%
{Joets88,Bisang98,Riecke98,Bisang99,toth02}. The most successful and widely
used method for the investigation of the complex convection patterns is
optical microscopy, based on the shadowgraph method \cite{Rasenat89}. It
utilizes the deflection of light rays in the spatially modulated director
field of the nematic. It has been applied to the determination of wave
vectors, onset thresholds and subcritical fluctuations of convection. One of
the problems encountered in this very efficient observation technique is the
complexity of the optics in the periodically deformed director field.
Simulations of the optical profiles have been presented by several authors
\cite{Rasenat89,Hirata82,Kondo83,Kosmo87,Joets86a,Joets86b,Amm98}, and the
consequences of in-plane director twist have been considered \cite%
{Grigutsch98,Amm99}. Although light propagation in such a two-dimensionally
inhomogeneous medium has been treated theoretically with different
approximation methods, and the qualitative relation between director
structure and observed intensity profile in the microscope is well
established, the method fails to provide quantitative access to deflection
amplitudes of the director field. With varying amplitudes of the spatially
periodic director modulations, both the positions of focal planes of the
patterns and the intensity profiles at given focal planes of the microscope
change in a complex way. Thus, the power of the method lies primarily in a
quantitative determination of the pattern wavelengths and the topology of
defect structures. Moreover, a fast (real-time) observation of the pattern
amplitude dynamics requires considerable bandwidth and signal processing
speed.

The wave vector spectrum can be determined qualitatively from a Fourier
transform of digitized microscopic transmission images \cite%
{Amm98,Amm97,Nagaya99}. A more efficient, quantitative way is the
quasi-optical Fourier transformation by means of laser diffraction. Laser
diffraction has been applied earlier to the study of Williams domains or
comparable dissipative patterns of nematics by Akahoshi et al.~\cite%
{Akahoshi}, Vistin and Yakovenko \cite{Vistin83} and Miike et al. \cite%
{Miike84}. In these studies, the evaluation of the scattering profile
remained to a large extend qualitative. A first experimental and theoretical
study of the laser diffraction efficiency of EHC has been presented by
Carroll \cite{Carroll72}, and Kashnow \cite{Kashnow73}. Scattering spots
designated to gratings in phase and amplitude of transmitted light can be
distinguished. They are generated by spatially modulated optical path and
light deflections, respectively.

In a structurally similar system, Bouvier and Scharf have employed the Jones
matrix method to calculate the diffraction efficiency of periodically
deformed director fields in cells with structured electrodes and compared it
to experimental data \cite{Scharf00}. Their method depends upon the
assumption of normal undeflected light transmission through the medium and
describes only the phase grating.

Comprehensive analysis of diffraction gratings formed by EHC has been
presented by Zenginoglou, Kosmopoulos and Papadopoulos \cite%
{Kosmo87,Zengi88,Zengi89,Zengi97,Papadopoulos99}. Various aspects of laser
diffraction by EHC have been considered, like the test of the validity of
geometrical optics \cite{Zengi97}, diffraction under oblique incidence \cite%
{Zengi88}, director oscillations and relaxation, and the dielectric regime
\cite{Papadopoulos99,Zengi01}. In a study of stochastically excited EHC \cite%
{johnPRE,johnPRL,behnbuch}, laser diffraction has been successfully applied
to characterize fundamental scaling laws in the statistical description of
pattern dynamics. The advantage of the laser diffraction technique over
shadowgraph observations is particularly evident in such an experiment where
data reduction is necessary to process pattern wavelengths and the
trajectory of the pattern amplitude in real time.

The previous study of fundamental scaling laws in stoch\-astically driven
EHC was based on several properties of the diffraction profiles which have
not been explicitly given there \cite{johnPRE,behnbuch}. This manuscript
deals with the underlying optical principles, it provides a justification of
the quantitative relations between diffraction profiles and director field
deflection amplitudes, and moreover, derives the absolute diffraction
efficiencies which allow the determination of not only growth rates and
relative pattern amplitudes but also the director tilt amplitudes in a
quantitative way. We demonstrate the application of the laser diffraction
technique to the study of amplitude dynamics of deterministically and
stochastically excited patterns. A rigorous treatment of light propagation
in two-dimensionally inhomogeneous director fields of EHC has not been
presented so far, therefore we will discuss the validity of several
approximations. We recollect the methods to calculate light propagation and
the corresponding phase and amplitude modulations in two-dimensionally
inhomogeneous director fields. A weakly nonlinear analytical calculation is
compared with numerical simulations of the full nonlinear equations and with
quantitative experimental data.

The paper is organized as follows: a short introduction into the basic
principles of the Carr-Helfrich mechanism and the involved dynamic equations
is given in the second section. We introduce the experimental setup and the
qualitative structure of the diffraction patterns in the third section. In
Sec. 4, we derive the analytical formula for the phase and amplitude
gratings generated by a weakly distorted director field, which also accounts
for first order effects of light deflection. In the course of the fourth
section, the quantitative numerical calculation of the diffraction profile
is performed. Although a great part of the equations derived in this section
have been communicated in earlier work by other authors, we consider it
helpful to include a comprehensive treatment of the optical background here,
in particular because in literature sometimes there seem to be contradictory
details of the calculations (see below). One obtains numerically the phase
and amplitude gratings produced by the mesogen layer for normal and oblique
incidence of monochromatic light. This allows quantitative predictions from
the combined effects of ray deflection and optical path length modulation.
We describe the diffraction efficiency of periodic nematic director
structures and compare our calculations with the approaches proposed in
literature. The numerical and analytical results are tested by comparison
with the experiment. Finally, we apply the method to the determination of
growth rates and Lyapunov exponents of pattern amplitudes in EHC and
demonstrate the power of the method to determine real-time amplitude
fluctuations of director field modes.

\section{Electroconvection}

Electrically driven convection in liquid crystals bases on the interaction
of free charges in the mesogen with external electric fields, and the
coupling of fluid flow to the deflection of the nematic director. A
comprehensive review is given, e.g., in Refs. \cite%
{Kramer,Kramer2,Kramer3,Kramer4}. The essential variables describing the
structures are the spatially modulated charge distribution $\tilde{q}(x,y,z)$
and the director tilt angle $\tilde{\varphi }(x,y,z)$, both are coupled via
the electrohydrodynamic equations. In an oscillating excitation field, the
two quantities have qualitatively different dynamic behaviour. For the
diffraction experiment as well as for the conventional shadowgraph images,
only the director field modulation is relevant and accessible.

\begin{figure}[ptb]
\begin{center}
\includegraphics[width=8.5cm]{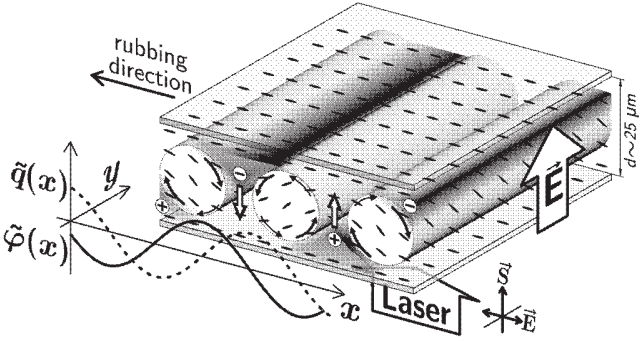}
\end{center}
\caption{ Schematic drawing of convection rolls and director field in a
nematic sandwich cell. A snapshot of the spatial modulations of director and
charge fields $(\tilde{\protect\varphi},\tilde{q})$ in the cell midplane is
sketched.}
\label{figehc}
\end{figure}

\begin{figure}[ptb]
\begin{center}
\includegraphics[width=8.5cm]{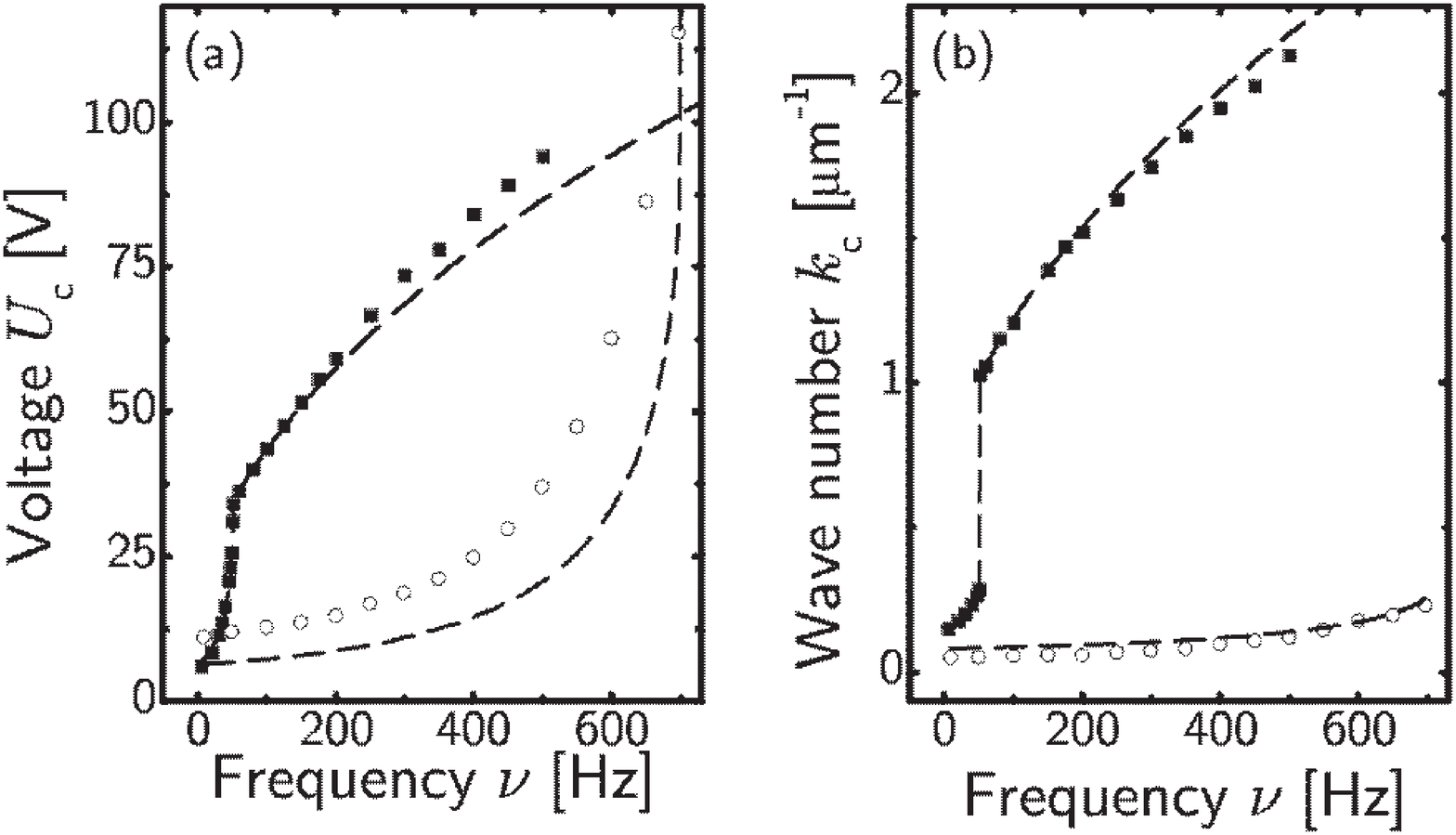}
\end{center}
\caption{ Stability diagram of the electroconvection patterns at the first
instability measured for cell 1 (25.8\thinspace\textmu{}m thick, $
\scriptscriptstyle\blacksquare$) and cell 2 (48.5\thinspace\textmu{}m thick,
$\circ$) under square wave excitation. The thresholds under sine wave
excitation are not much different from the square wave case (at comparable
effective voltages). Lines are from the analytical calculation, where known
material parameter from independent experiments, have been used, if
available. The remaining unknown parameter have been obtained from the fit
to the experimental data. The cut off frequency $\protect\nu_{\text{c}}=51$%
\thinspace Hz for cell 1 separates conduction and dielectric regimes. In
case of cell 2 the much higher cut-off frequency is outside of the presented
range. }
\label{figthresholds}
\end{figure}

A sketch of the experimental geometry is given in Fig.~\ref{figehc}. The
nematic director alignment at the glass plates is fixed along $x$ by surface
treatment. The ground state is a uniform director field in the cell. When an
electrical field $E=U/d$ is applied between the transparent ITO-electrodes
at the glass plates, free charge carriers (ionic impurities or dopants) in
the nematic fluid are accelerated and initiate a macroscopic flow. The
conductivity anisotropy of the material in combination with small
fluctuation modes of the director tilt lead to lateral charge separation in $%
x$\thinspace direction and a periodically modulated flow field, which in
turn couples to the director field by hydrodynamic equations. At the
critical field $E_{\text{c}}$, stabilizing elastic and dielectric torques on
the director are outmatched by destabilizing hydrodynamic torques. Under
standard conditions, the system exhibits a forward bifurcation to normal
rolls (wave vector along the easy axis of the director) or to oblique rolls.
Threshold voltage and critical wave number are frequency dependent. The
pattern stability diagram of the two sandwich cells studied here is shown in
Fig.~\ref{figthresholds}. The nematic material is \emph{Mischung 5}, a
mixture of four di\-substituted phenyl-benzoates \cite{johnPRE}, material
parameters in Table I. The first cell has been prepared with the pure,
undoped material, which has a low conductivity and correspondingly low
cut-off.

The nematic mixture in sample 2 has been doped with 0.5 mass%
\textperthousand{} tetrabutyl-ammonium bromide. Therefore its conductivities
are much higher and the increased amount of charge carriers shifts the
cut-off frequency out of the frequency range of measurements. It leads to a
much more stable pattern amplitude characteristics near the threshold. In
the undoped material, the content of charge carriers is comparably small
('natural' impurities after synthesis) and in the experiment, the threshold
voltage is subject to certain small but measurable long-term fluctuations.

In Fig. \ref{figthresholds}, one distinguishes the low frequency
'conduction' regime and the high frequency 'dielectric' regime of cell 1,
with a distinct jump in the wave number at the cut off frequency. Only the
low frequency regime is in the accessible frequency range in cell 2. All
measurements in this study are performed in the conduction regime, where the
director field performs only moderate oscillations synchronous with the
excitation frequency, but keeps its sign during the field cycles. However,
there is no principal limitation for an application of the presented setup
to structures of the dielectric regime \cite%
{Papadopoulos99,Zengi01,Bohatsch99}.

The mathematical description is based on the Maxwell and Navier Stokes
equations. The linear stability analysis of the torque balance uses a test
mode ansatz for director deflections and charge density modulations
\begin{align}
\tilde{\varphi}(x,z,t) & =\varphi_{t}\cos(k_{x}x)\cos(k_{z}z),
\label{testmodeansatz1} \\
\tilde{q}(x,z,t) & =q_{t}\sin(k_{x}x)\cos(k_{z}z),   \label{testmodeansatz2}
\end{align}
where $k_{x}$ is the periodicity of the pattern. Because the director is
fixed parallel at the glass plates, the boundary conditions $\varphi(\pm
z=d/2)=0$ enforce $k_{z}=(2n+1)\pi/d$, $n$ integer. Near onset, we consider
only the ground mode $k_{z}=\pi/d$. In the investigated parameter regions
here only normal rolls appear, and no $y$ dependence has to be considered.
Linearization leads to a linear ordinary differential equation system in $%
(q,\varphi)$. The solution for a constant electric field amplitude $E$
involves a $2\times2$ non-symmetric time evolution matrix $\mathbf{T}$ \cite%
{Behn98} which depends on the wave vector $k_{x}$ of the particular test
mode
\begin{equation}
\binom{q}{\varphi}(t)=\mathbf{T}(E,k_{x},t)\binom{q}{\varphi}(0).
\label{qphionestep}
\end{equation}
At square wave excitation, where only the sign of $E$ alternates, the time
evolution at points with alternating sign of $E$ is given by a product of
matrices with constant coefficients
\begin{align}
\binom{q}{\varphi}(t_{n}) & =\mathbf{T}^{\pm}(\Delta t_{n})\cdots \mathbf{T}%
^{-}(\Delta t_{1})\mathbf{T}^{+}(\Delta t_{0})\binom{q}{\varphi }(0),
\label{growthrate1} \\
t_{n} & =\sum\nolimits_{i=0}^{n}\Delta t_{i},\ E(t)=\pm E,
\end{align}
where $\Delta t_{i}$ are the time intervals between consecutive jumps. The
solution at intermediate times is calculated with Eq. (\ref{qphionestep}).
The involved material parameters are listed in Table 1, further details are
given in \cite{Behn98}.

The largest of the two real eigenvalues of the matrix product in Eq. (\ref%
{growthrate1}) is related to the Lyapunov exponent of the system \cite%
{Behn98} and describes the asymptotic behaviour of a small initial
perturbation in $(q,\varphi)$. In case of a periodic excitation, all $\Delta
t_{i}=\Delta t=1/(2\nu)$ are equal and after $n$ full periods of the ac
excitation, the product in Eq. (\ref{growthrate1}) can be split in repeated
blocks of the product $\mathbf{T}^{+}\mathbf{T}^{-}$

\begin{equation}
\binom{q}{\varphi}(t_{n})=(\mathbf{T}^{+}\mathbf{T}^{-})^{n/2}\binom {q}{%
\varphi}(0).   \label{qpsisquare}
\end{equation}
For $t_{n}\gg\Delta t$, the amplitudes of both variables grow or decay
exponentially, and the (dimensionless) largest real eigenvalue $\lambda
_{1}(E,k_{x},\nu)$ of $\mathbf{T}^{+}\mathbf{T}^{-}$ gives the growth or
decay rate $\nu\lambda_{1}$. The maximum $\lambda_{1}(E,k_{x},\nu)$ of all $%
k_{x}$ selects the critical wave number $k_{\text{c}}$. The theoretical
threshold field $E_{\text{c}}$ at a given frequency $\nu$ is determined by
the first positive value of $\lambda_{1}(E,k_{\text{c}},\nu)$ with
increasing $E$. This value coincides with results from the Floquet theorem.

With any optical detection methods (shadow graph or diffraction), only $%
\varphi_{t}$ is observable and asymptotically for $t\gg\Delta t$, (in the
limit of small $\varphi_t$)
\begin{equation}
\varphi_{t}=\varphi_{0}\text{e}^{\nu\lambda_{1}\,t},   \label{lyapunov1}
\end{equation}
where, $\varphi_{0}$ is the initial amplitude of the considered mode,
related to fluctuations of director and charge fields. For large amplitudes,
$\varphi_{t}$ is limited by nonlinearities that will not be considered in
the linear treatment. For sample 1, the theoretical threshold curves $U_{%
\text{c}}(\nu)=E_{\text{c}}d$ and $k_{\text{c}}(\nu)$ can be fitted to the
experimental data over a wide frequency range with most of the involved
material parameters taken from independent experiments, see Table I. Some
remaining unknown viscoelastic parameters are obtained from the fit. With
this completed set of parameters, the theoretical $\lambda_{1}(E,k_{\text{c}%
},\nu)$ dependence can be calculated analytically as a function of the
excitation frequency $\nu$.

\section{Laser diffraction experiment}

The diffraction of laser light by the director pattern provides the
opportunity to analyse the spatial mode spectrum in real time. Figure \ref%
{figsetup} sketches the experimental setup consisting of a low power ($%
\approx$1 mW) He-Ne laser, the liquid crystal (LC) cell mounted in a Linkam
microscope hot stage TMS 600, a photodiode or alternatively a diffusely
reflecting screen for 2D camera images.

The photodiode can be moved by a stepper motor in horizontal $x$\thinspace
direction across the scattering image. Its aperture is 3~mm$\times$3~mm. At
a distance $\ell$ of approx 800 mm from the LC cell, this corresponds to an
angular resolution of 5 mrad. The 2D images of the CCD camera are used for
the qualitative characterization of the diffraction patterns only (see
appendix), while all quantitative intensity measurements are performed with
the photodiode.
\begin{figure}[ptb]
\begin{center}
\includegraphics[width=8.5cm]{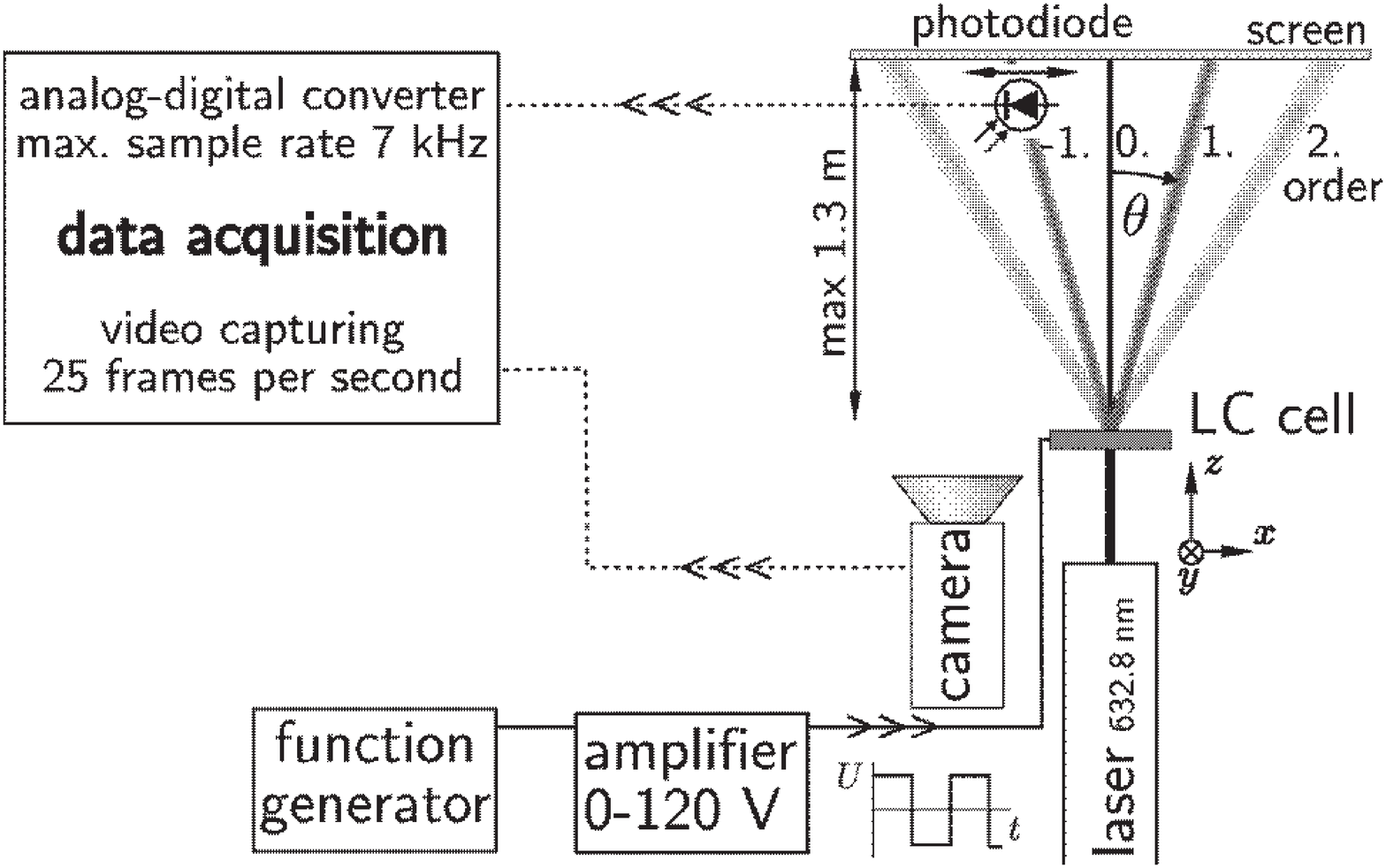}
\end{center}
\caption{Sketch of the experimental setup. The driving voltages are
synthesized by a computer with digital analog (DA) converter and analog
amplifier. The photodiode signal is sampled by the computer at a maximum
rate of 7~kHz with an accuracy of 12 bit, or alternatively sampled by a
digital voltmeter at a rate of 50 Hz with an accuracy of at least 6 digits. }
\label{figsetup}
\end{figure}

The test modes of Eqs.~(\ref{testmodeansatz1},\ref{testmodeansatz2})
correspond to a one-dimen\-sional stripe pattern along $x$ in the microscope
image. For the sample cell 1 studied here, it appears at intermediate
frequencies, from the Lifshiz-point (below 20 Hz) to the cut-off frequency ($%
\approx50$~Hz), as the first instability. At lower frequencies, the wave
vector at onset has a non-zero $y$ component.
\begin{figure}[ptb]
\begin{center}
\includegraphics[width=8.5cm]{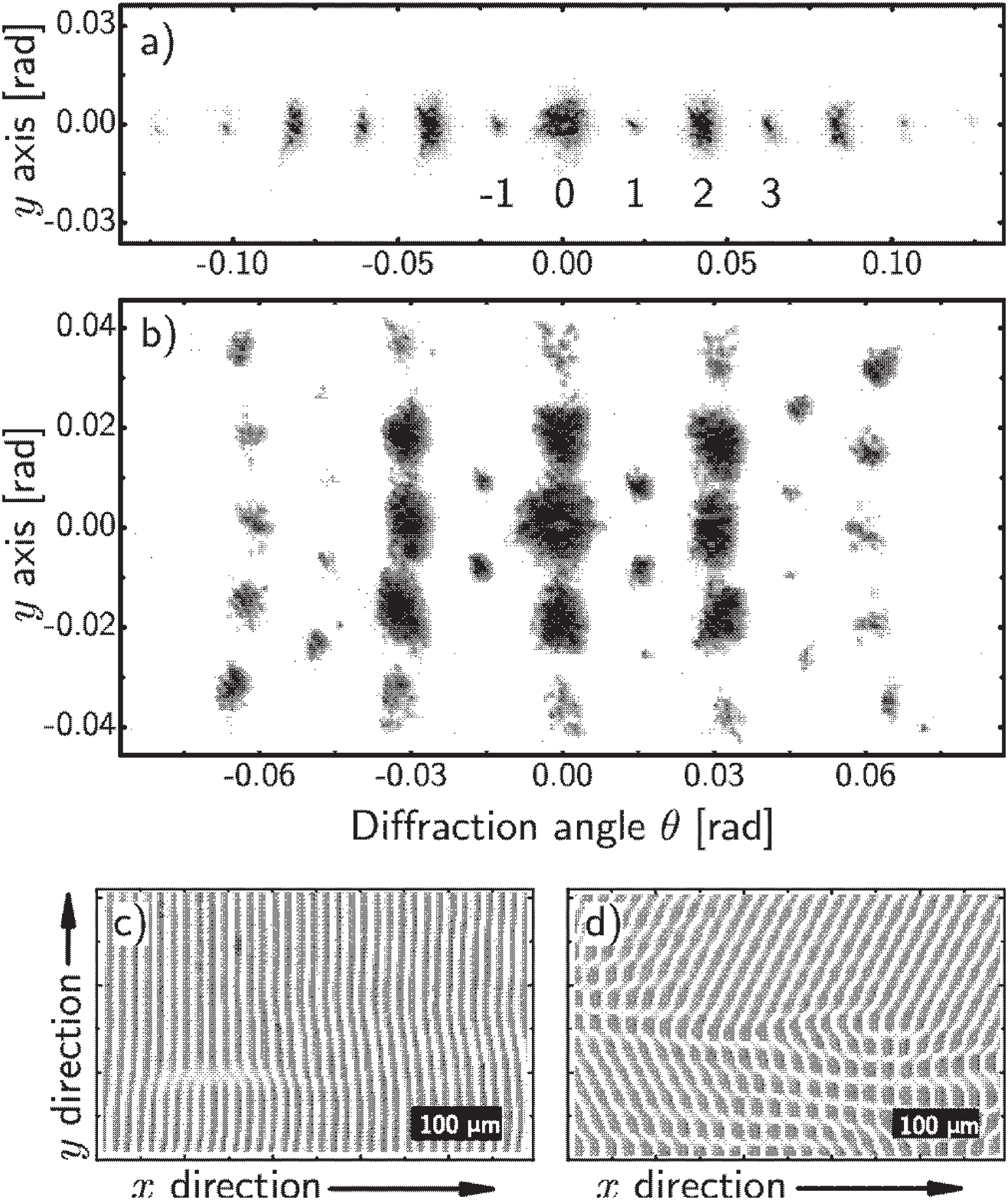}
\end{center}
\caption{ Snapshots of diffraction images (a,b) and respective microscope
images (c,d) for normal rolls at 30 Hz (a,c) and oblique rolls at 10 Hz
(b,d), cell 1. Numbers in (a) mark the diffraction order. In the lower left
part in (c), a localized defect of the roll pattern is visible. The cell
thickness is 25.8 \textmu{}m.}
\label{figsampleimages}
\end{figure}

Figure~\ref{figsampleimages} demonstrates the two cases of normal and
oblique rolls for square wave excitation at 30\thinspace Hz and 10
\thinspace Hz, resp., at voltages slightly above the convection threshold.
In such cases where one or two superimposed wave vectors form the spatial
structure, the diffraction experiment provides the pattern wave lengths,
orientations and amplitudes. Defects and domain sizes will primarily
influence width and fine structure of individual diffraction peaks but are
not directly accessible from the profiles.

\section{Optics}

\subsection{Light propagation}

The light intensity at the position of the diffraction reflexes is directly
related to the amplitude of the spatial director mode. The problem of light
propagation in a periodically modulated director field has been investigated
by different groups before \cite%
{Kondo83,Kosmo87,Carroll72,Kashnow73,Scharf00}. These previous works can be
grouped into two different approaches. One is based on the solution of the
Maxwell equations, the calculation of the spatial distribution of $\vec{E}$
and $\vec{D}$. Yet, a complete solution of the Maxwell equation with
boundary conditions can only be obtained numerically for the present
problem. A linearized wave propagation approach can be found, for example,
in Ref. \cite{Zengi97}. The alternative method is based on the calculation
of light ray's using the eikonal method \cite{Trainoff02} or crystal optical
methods \cite{Kosmo87,Zengi89}. Here, we give a short outline of the
calculation of phase and amplitude of the laser light passing the LC layer
using crystal optics.

The polarization vector $\vec{E}$ of the incident laser light is adjusted
along the director easy axis, since only the extraordinary wave is relevant
for the diffraction effect (see Fig.~\ref{figehc}). Ordinary rays pass the
LC layer without deflections and phase modulation.

\begin{figure}[tbh]
\begin{center}
\includegraphics[width=8.5cm]{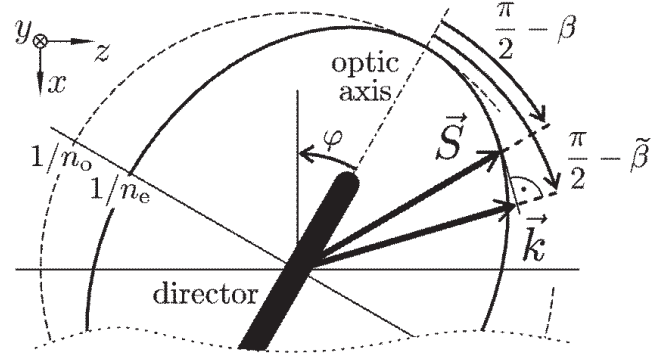}
\end{center}
\caption{ Definition of angles on the ray ellipsoid, $\protect\varphi$~:\
director deflection, $\protect\pi/2-\protect\beta$~:~angle between optic
axis and {Poynting-}vector $\vec{S},$ $\protect\pi/2-\protect\tilde{\beta}$%
~:~angle between optic axis and normal of plane wavefront $\vec{k}.$}
\label{defangle}
\end{figure}

In the lowest order of approximation, we may assume a straight propagation
of the electromagnetic wave without any deflection of the light beam, $r(z;x_0)$ being the $x$%
-coordinate of the ray
\begin{gather}
r^{\prime }(z;x_0)=\text{d}r(z;x_0)/\text{d}z=0 \\ r(z;x_0)=x_{0}.
\end{gather}
It yields no amplitude modulation but a first estimation of the phase profile
of the light that has penetrated the cell. Under these assumption the phase
$\psi _{\text{s}}(x)$ in dependence on the amplitude of director deflections is
given by (see Fig.~\ref{defangle})
\begin{gather}
\psi _{\text{s}}(x)=k_{\text{{\tiny L}}}\int\limits_{-d/2}^{d/2}\!\!\!n_{%
\text{eff}}^\text{p}(r,r^\prime,z)\,\text{d}z=k_{\text{{\tiny L}}}n_{\text{e}%
}d+\Delta \psi _{\text{s}}(x), \\
n_{\text{eff}}^{\text{p}}(\tilde{\beta})=\frac{n_{\text{o}%
}n_{\text{e}}}{\sqrt{n_{\text{o}}^{2}\cos ^{2}\tilde{\beta}
+n_{\text{e}}^{2}\sin ^{2}\tilde{\beta}}}, \\
\tilde{\beta}(r,r^{\prime },z)=\varphi_t \cos(k_x r)\cos(k_z z), \\
\Delta \psi _{\text{s}}(x_{0})\approx -k_{\text{{\tiny L}}}n_{\text{e}}d%
\frac{n_{\text{e}}^{2}-n_{\text{o}}^{2}}{8n_{\text{o}}^{2}}\,\varphi
_{t}^{2}\cos (2k_{x}x_{0}),  \label{phasestraight}
\end{gather}%
where $n_{\text{eff}}^{\text{p}}$ is the effective refractive index and $n_{%
\text{o}},n_{\text{e}}$ are the ordinary and the extraordinary refractive
indices of the nematic material (see Table 1). Only the first non constant
term in the series expansion of $n_{\text{eff}}^{\text{p}}$ enters the
result, the constant phase does not contribute to the diffraction profile.

The exact calculation of light propagation basing on the calculation of ray
paths $(r(z;x_{0}),z)$ uses the \textsc{Fermat} principle. From the symmetry of
the problem, no change of polarization of the light can occur. If the wave
number of the laser light $k_{\text{\tiny L}}$ is much larger than the wave
number of the pattern, the \textsc{Fermat} principle can be applied in the
birefringent material. The minimum condition is
\begin{gather}
\int\limits_{-d/2}^{d/2}\!\!\!n_{\text{eff}}^{\text{r}}(r,r^{\prime
},z)\,\text{d}s=\min ,  \label{fermat} \\ n_{\text{eff}}^{\text{r}}(r,r^{\prime
},z)=\sqrt{n_{\text{e}}^{2}\cos ^{2}\beta +n_{\text{o}}^{2}\sin ^{2}\beta },\\
\beta(r,r^{\prime },z)=\varphi_t \cos(k_x r)\cos(k_z z)-\arctan r^{\prime},\\
\text{d}s=\sqrt{1+r^{\prime 2}}\text{d}z,
\end{gather}%
where $n_{\text{eff}}^{\text{r}}$ is the effective ray index \footnote{%
In literature, e.g. \cite{Rasenat89,Rehberg91a,Richter92}, there has sometimes
been a confusion about the appropriate usage of the ray index or refractive
index in the computation of light propagation through a birefringent nematic
layer with periodically deformed director field.}. Applying the
\textsc{Euler-Lagrange} formalism leads to a ordinary differential equation of
second order in the displacement $r(z;x_0)$ of a ray
\begin{gather}
r^{\prime \prime }=\varphi _{t}\frac{n_{\text{e}}^{2}-n_{\text{o}}^{2}}{%
n_{\text{e}}^{2}n_{\text{o}}^{2}}\left( 1+r^{\prime }\right) \left(
t_{1}+t_{2}\right) ,  \label{fullEulerDGL} \\ t_{1}=k_{z}\cos (k_{x}r)\sin
(k_{z}z)\left( n_{\text{e}}^{2}\cos ^{4}\beta -n_{\text{e}}^{2}r^{\prime }\cos
^{3}\beta \sin \beta -\right. \notag \\
~~~\left. -n_{\text{o}}^{2}r^{\prime }\cos \beta \sin ^{3}\beta -n_{\text{%
o}}^{2}\sin ^{4}\beta \right) ,  \notag \\ t_{2}=k_{x}\sin (k_{x}r)\cos
(k_{z}z)\left( n_{\text{e}}^{2}r^{\prime }\cos ^{4}\beta +n_{\text{e}}^{2}\cos
^{3}\beta \sin \beta +\right. \notag \\
~~~\left. +n_{\text{o}}^{2}\cos \beta \sin ^{3}\beta -n_{\text{o}%
}^{2}r^{\prime }\sin ^{4}\beta \right) .  \notag
\end{gather}%
Eq. (\ref{fullEulerDGL}) is an exact result which we use in a \textsc{%
Runge-Kutta} algorithm for a numerical computation of the ray paths.
Expanding of (\ref{fermat}) to first order in the amplitude $\varphi _{t}$
of the spatial mode of director deflection yields
\begin{equation}
r^{\prime \prime }\approx \frac{n_{\text{e}}^{2}-n_{\text{o}}^{2}}{n_{%
\text{o}}^{2}}\varphi _{t}k_{z}\cos (k_{x}r)\sin (k_{z}z). \label{deflectdgl}
\end{equation}%
The integration of Eq.~(\ref{deflectdgl}) with the initial values
\begin{equation}
r|_{z=-d/2}=x_{0},\, r^{\prime }|_{z=-d/2}=0
\end{equation}
and the approximation $\cos (k_{x}r)=\cos (k_{x}x_{0})$ leads to a path which
enters at $x_{0}$ and
\begin{equation}
r(z;x_{0})=x_{0}-\frac{1}{k_{z}}\frac{n_{\text{e}}^{2}-n_{\text{o}}^{2}}{n_{%
\text{o}}^{2}}\varphi _{t}(1+\sin k_{z}z)\cos k_{x}x_{0}.  \label{rays}
\end{equation}%
To describe the focussing effect in the shadowgraph method the second order
term in $\varphi _{t}$ must be considered, it can be found in Ref.~\cite%
{Rasenat89}. For the diffraction pattern, mainly the phase is important (at
least for small director deflections) and therefore no consideration of
higher order terms are necessary. In order to calculate the resulting phase
beyond the cell, we have to distinguish between the direction of propagation
of energy flux $\vec{S}$ and the normal of wavefronts $\vec{k}$ (Fig. \ref%
{defangle}). In uniaxial birefringent material the relation is \cite%
{Encyclop}%
\begin{equation}
\tan \left( \frac{\pi }{2}-\tilde{\beta}\right) =\frac{n_{\text{e}}^{2}}{%
n_{\text{o}}^{2}}\tan \left( \frac{\pi }{2}-\beta \right) .
\end{equation}
The phase can be calculated using different indices
\begin{equation}
\frac{\psi}{k_{\text{\tiny L}}} =\int
n_{\text{eff}}^{\text{r}}\text{d}s=\int
n_{\text{eff}}^{\text{p}}\text{d}|\vec{k}|=\int
n_{\text{eff}}^{\text{p}}\cos(\beta-\tilde{\beta})\text{d}s .
\end{equation}%
Finally, the calculated lateral phase difference along curved paths at the exit
position for small director deflections is basically the same as the
result for straight transmission (\ref{phasestraight})%
\begin{equation}
\Delta \psi (r,z)|_{z=d/2}=\Delta \psi _{\text{s}}(x_0). \label{psivsphi}
\end{equation}%
This result is in agreement with the relation in \cite{Trainoff02} calculated
from the eikonal equation. Also, the next order term in $\varphi _{t}^{3}$ is
stated in \cite{Trainoff02}. The consideration of the displacement
$r(z;x_{0})-x_{0}$ gives
\begin{align}
\Delta \psi (\tilde{x},z)|_{z=d/2}& \approx -k_{\text{{\tiny L}}}n_{\text{e}%
}d\frac{n_{\text{e}}^{2}-n_{\text{o}}^{2}}{8n_{\text{o}}^{2}}\,\varphi
_{t}^{2}\cos (2k_{x}\tilde{x}), \\ \tilde{x}& =r(d/2;x_{0}),
\end{align}%
$\tilde{x}$ denotes the exit position $r(d/2;x_{0})$ of a beam entering at $%
x_{0}$ which may differ from $x_{0}$ due to the ray deflection. Therefore,
the first order bend of the light paths is an effect of the birefringence
(the inclination of the optic axis of the nematic material respective to the
incident beam direction) and is not the result of the periodic modulation of
the refraction (connected with the ray index) in $x$-direction.

The displacement of the rays after propagating the cell and the conservation
of energy gives the intensity $\mathcal{I}$ in terms of the incident
intensity $\mathcal{I}_0$, and the amplitude of the electric field $\mathcal{%
E}$ (with $\mathcal{E}^{2}\propto\mathcal{I}$) in the exit plane
\begin{gather}
\mathcal{I}_{0}\delta x_{0}=\mathcal{I}\delta\tilde{x},  \label{intensansatz}
\\
\mathcal{I}_{0}\text{d}x_{0}=\mathcal{I}\left. \left[ r(z;x_{0}+\text{d}%
x_{0})-r(z;x_{0})\right] \right\vert _{z=d/2}, \\
\frac{\mathcal{I}}{\mathcal{I}_{0}}=\frac{\mathcal{E}^{2}(\tilde{x})}{%
\mathcal{E}_{0}^{2}}=\left(1+2\frac{k_{x}}{k_{z}}\frac{n_{\text{e}}^{2}-n_{%
\text{o}}^{2}}{n_{\text{o}}^{2}}\varphi_{t}\sin k_{x}x\right)^{-1}.
\label{intensfinal}
\end{gather}
The ansatz (\ref{deflectdgl}) is only correct in the limit of small $%
\varphi_{t}$ when rays leaving the cell in parallel in good approximation. A
correction resulting from more exact treatment based on $\nabla\vec{S}=0$ can
be found in Ref.~\cite{Kosmo87}.

\begin{figure}[ptb]
\begin{center}
\includegraphics[width=8.5cm]{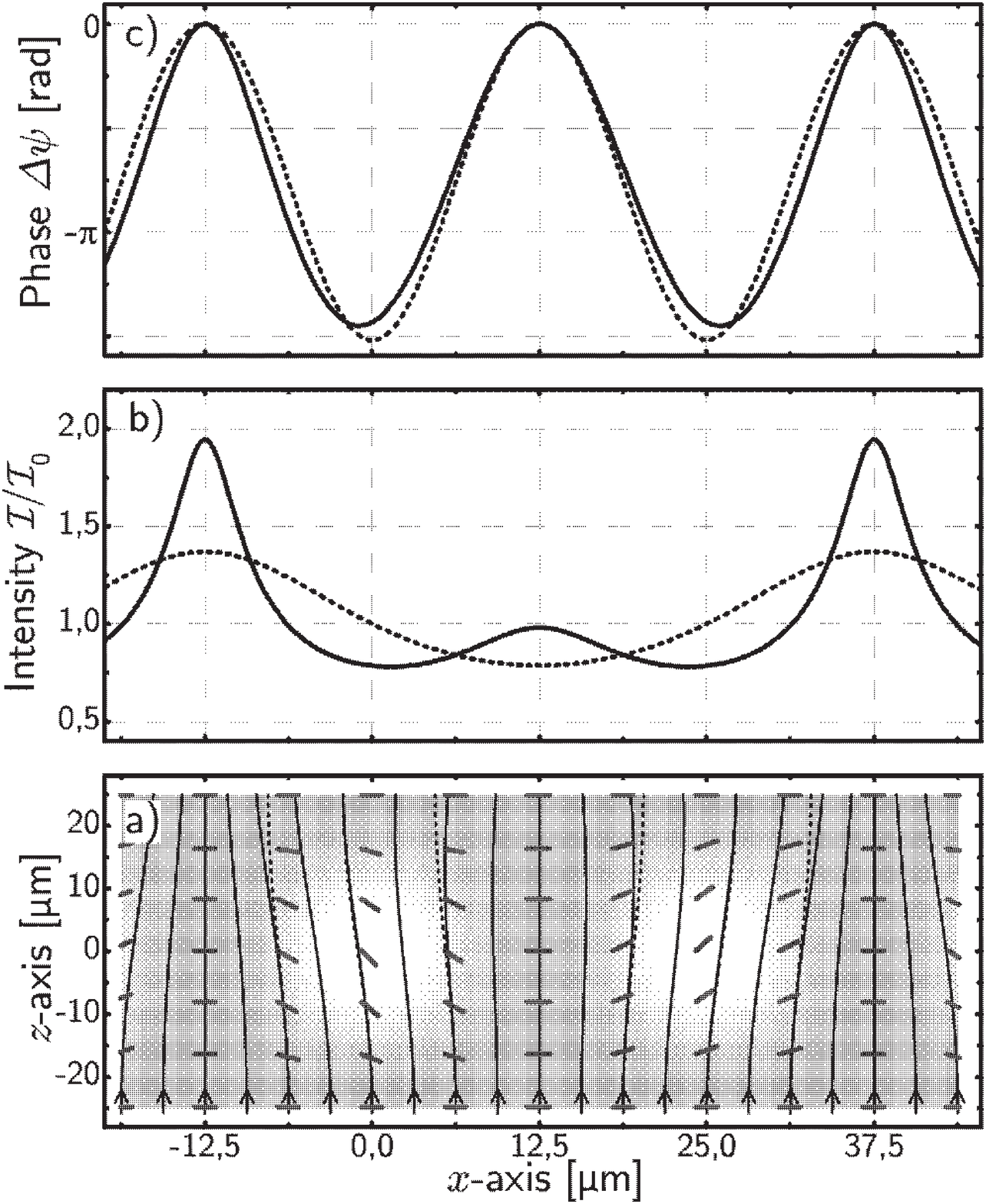}
\end{center}
\caption{Numerically (solid) and analytically (dotted) calculated ray
propagation (a), intensity (b), and relative phase (c) of light after
penetrating the nematic layer at $z=25$~\thinspace\textmu m. The assumed
director tilt ($\protect\varphi=20$\textdegree{}) is schematically depicted in
(a), the grey scale visualizes the effective refraction index for straight
light propagation. The periodicity of the amplitude grating is that of the
director field. In contrast, the dominating phase modulation has
twice the wave number of the director field. A cell thickness of 50 \textmu{}%
m and optical parameters of Tab. I have been assumed. }
\label{figraysvertical}
\end{figure}

Figure \ref{figraysvertical}a) illustrates the light propagation from
different entry positions along the $x$-axis in the case of a strong
director tilt amplitude $\varphi=20$\textdegree{}. It compares the
numerically calculated paths with the analytical result from (\ref{rays}).
Also the intensity and the phase at exit position are depicted in Fig \ref%
{figraysvertical}b,c). The analytical result for the phase differs hardly
from the exact numerically calculation whereas the first order approximation
in intensity differs even qualitatively from the numerical values.
Fortunately, the diffraction efficiency is dominated by the phase profile
and therefore the conclusions drawn from diffraction profiles about director
deflections are correct up to large deflections angles.

\subsection{Diffraction profiles}

The cell is illuminated with a normally incident planar wave, polarized in $x
$ direction. The area contributing to diffraction is a circular spot with
radius $s\approx 0.5$\thinspace mm. In general, the electric field at the
rear of the cell can be written as
\begin{equation}
\mathcal{\widehat{E}}|_{z=d/2}=\mathcal{E}(x)e^{\text{i}\Delta\psi(x)},
\end{equation}
with the amplitude $\mathcal{E}(x)$ and $\Delta\psi(x)$ the lateral phase
difference of the wave at the position $(x,z=d/2)$. For a one-dimensional
modulation (Eqs. (\ref{testmodeansatz1},\ref{testmodeansatz2})), each
location $x$ can be considered as the origin of a spherical wave. The
diffraction intensity at $\ell \gg s$ into the angle $\theta$ in the ($x,z$%
)\thinspace plane is
\begin{align}
\text{d}E(x,\vec{k}_{\text{{\tiny L}}};\vec{l}) & =\frac{\mathcal{E}(x)}{\ell%
}\text{e}^{\text{i}\left[ -\vec{k}_{\text{{\tiny L}}}\vec{l}+\Delta \psi(x)%
\right] }\text{d}x\text{d}y, \\
\vec{k}_{\text{{\tiny L}}}\vec{l} & =k_{\text{{\tiny L}}}\ell-xk_{\text{%
{\tiny L}}}\sin\theta,
\end{align}
where $k_{\text{{\tiny L}}}$ is the wave number of the incident light,
the vector $\vec{l}$ connects the cell with the detector position and $\ell$
the distance between cell and detector.
Integration over a circular area with radius $s$ of the
illumination spot gives the amplitude of the complex wave with the
corresponding intensity
\begin{gather}
E(\theta)\propto\int\limits_{-s}^{s}\int\limits_{-\sqrt{s^{2}-y^{2}}}^{\sqrt{%
s^{2}-y^{2}}}\!\!\!\!\!\!\mathcal{E}(x)\text{e}^{\text{i}\,\left[ x k_{\text{%
{\tiny L}}}\sin\theta+\Delta\psi(x)\right] }\text{d}x\text{d}y,
\label{FieldAmplitude} \\ I(\theta)\propto|E(\theta)|^{2}.
\end{gather}

To calculate the complete diffraction function numerically, we consider both
the spatial modulation of $\mathcal{E}(x)$ (amplitude grating) as well as
the phase modulation $\Delta\psi(x)$ (phase grating). Basically, the first
one is effective for shadow graph images. In contrast, when the director
modulation is small, the latter plays the dominating role in the diffraction
characteristics. Therefore, we will consider for analytical approximations
the phase grating alone.

In case of small director deflections, the differences between exit position
$\tilde{x}$ and entry point $x_{0}$ can be neglected and the periodicity of
the phase grating can be written as
\begin{equation}
\Delta\psi(x)=\Delta\psi_{\max}\cos(2k_{x}x).   \label{psidiffraction}
\end{equation}
It is twice that of the director field and therefor diffraction reflexes
from the the phase grating appear only at even order $n$. On the other hand
the periodicity of the amplitude grating (\ref{intensfinal}) is the same of
the director field and it contributes also on odd order reflexes.

The integration of (\ref{FieldAmplitude}) gives the intensities $%
I_{n}=I(\theta_{n})$ of the $n-$th order diffraction spots. With the
assumptions $\mathcal{E}(x)=$const. and (\ref{psidiffraction}) the
intensities at angles $\theta_{n}$ are described by Bessel functions $J_{n/2}
$ with the amplitude of the laser light phase modulation in the argument
\begin{equation}
\frac{I_{n}(\varphi_{t})}{I_{0}}=\frac{|E(\theta_{n})|^{2}}{|E(\theta
_{0})|^{2}}=J_{n/2}^{2}\left( \Delta\psi_{\max}\right) \approx\frac {%
\Delta\psi_{\max}^{n}}{2^{n}[(n/2)!]^{2}},
\end{equation}
where $I_{0}$ is the intensity of the mean beam at the ground state. Using (%
\ref{psivsphi}) and (\ref{phasestraight}) gives the final quantitative
relation between normalized diffraction intensity and director tilt
amplitude at the dominating second order reflex
\begin{equation}
\frac{I_{2}(\varphi_{t})}{I_{0}}=\frac{1}{4}\left[ k_{\text{{\tiny L}}}n_{%
\text{e}}d\frac{n_{\text{e}}^{2}-n_{\text{o}}^{2}}{8n_{\text{o}}^{2}}\right]
^{2}\varphi_{t}^{4}\text{, where }I_{2}\ll I_{0}.   \label{IvsPhi}
\end{equation}

The numerically obtained diffraction efficiencies in the limit of small $%
\varphi_{t}$ confirm the relation (\ref{IvsPhi}) including the prefactor.
The analytical approximations are satisfactory up to $\varphi\sim$30%
\textdegree{} for thin cells.

The numerical calculation offers an easy way to consider effects from
oblique illuminations, an important point to understand the sensitivity of
the diffraction images to a non perfect sample orientation. A detailed
discussion is found in the appendix.

\subsection{Experimental test \label{secexperimentaltest}}

We tested the presented calculations with the setup sketch\-ed in
Fig.\thinspace \ref{figsetup} with sample 2. The temperature is stabilized at
32\thinspace \textcelsius{} and the measurement is performed at 500\thinspace
Hz sine excitation. Due the high conductivity the cut-off frequency is shifted
above 600 Hz and the pattern is more permanent at 500~Hz sinodial excitation
than in cell 1. A circular aperture with 0.5\thinspace mm in dia\-meter defines
the illumination spot. A smaller movable photo diode (0.3\thinspace$\times$
0.6\thinspace mm) is positioned in a distance of $\ell=820$ mm, to obtain a
high resolution in diffraction angle. The analog-digital-converter is a
programmable Kethley multimeter with a resolution of 24 bit and a sampling rate
of 50\thinspace s$^{-1}$.

\begin{figure}[ptb]
\begin{center}
\includegraphics[width=8.5cm]{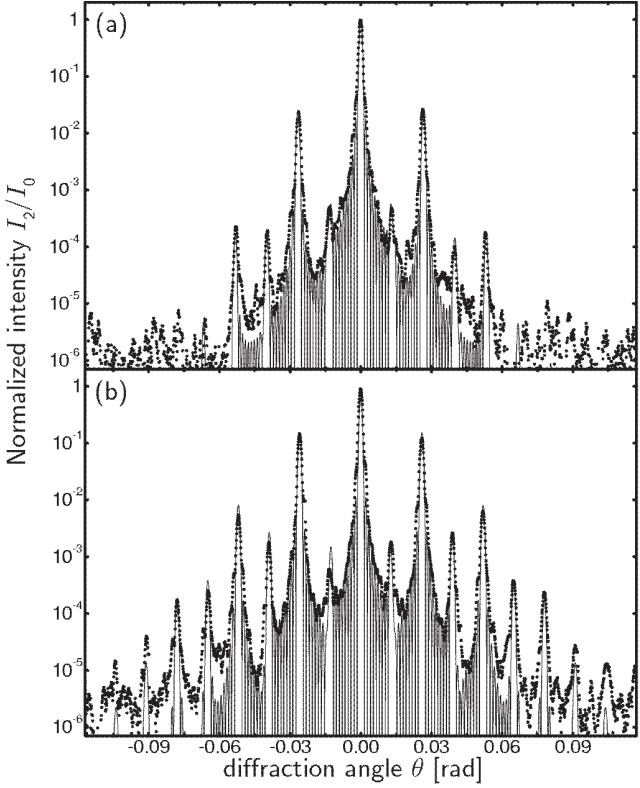}
\end{center}
\caption{ Comparison of numerical calculated diffraction profiles (solid
line) and measured intensities (dots), normalized to the primary beam
intensity. The measured profile at $U=28.79$\thinspace V (a) and $U=28.90$%
\thinspace V (b) corresponds to the calculation with amplitudes $\protect%
\varphi=7.4$\textdegree{} (a) and $\protect\varphi=12.6$\textdegree{} (b).
The corresponding deflections from measured $I_{2}/I_{0}$ using (\protect\ref%
{IvsPhi}) are 7.5\textdegree{} and 11.0\textdegree{}. The pattern wave
length $\protect\lambda_{\text{dir}}=48$\thinspace\textmu m, cell thickness
48.5 \textmu{}m, $U_{\text{c}}=28.73$\thinspace V. }
\label{figlaserdiffractionsim}
\end{figure}

Figure\thinspace\ref{figlaserdiffractionsim} shows the baseline corrected
intensity along $x$~direction in the $x$-$z$ plane ($y=0$ in Fig. \ref%
{figsampleimages}) for two applied voltages slightly above the instability
threshold. All profiles are normalized with the transmitted primary beam
intensity at $\theta =0$\thinspace. Note the logarithmic intensity scale. In
order to discriminate the diffracted light from a small constant background
(scattered light from glass plates and small amplifier offset), we subtract
the constant signal of the order of 10$^{-4}$, detected at large deflection
angles ($\theta >0.5$\thinspace rad), from all measurements. In
superposition with the constant offset, small fluctuations in order of 10$%
^{-7}$ are observed which drop below the detection level when the sample
temperature is increased above 80\thinspace\textcelsius{} into the isotropic
phase. The intensity profiles are compared with numerical calculations for
different director amplitudes. The best fits lead to director deflections $%
\varphi=(7.4\pm0.3)$\textdegree{} and $\varphi=(12.6\pm0.3)$\textdegree{}, $%
\epsilon=U/U_{\text{c}}-1\approx2\times10^{3}$ and $\epsilon\approx5%
\times10^{3}$, resp. (Fig.\thinspace \ref{figlaserdiffractionsim}). The
normalized intensity at the second order reflex coincides well with the
value obtained from analytical treatment (Eq. (\ref{IvsPhi})).

\section{Applications}

\subsection{Study of EHC near the threshold \label{secbifurcation}}

The sensitivity in the detection of small pattern amplitudes and the
quantitative relation between diffraction intensity and director tilt
provides the opportunity to study electroconvection near the onset threshold
experimentally. Amplification of thermal fluctuations slightly below the
threshold has been studied previously with the shadowgraph method \cite%
{Bisang98,Bisang99,Rehberg91,scherer00,scherer02}.

\begin{figure}[ptb]
\begin{center}
\includegraphics[width=8.5cm]{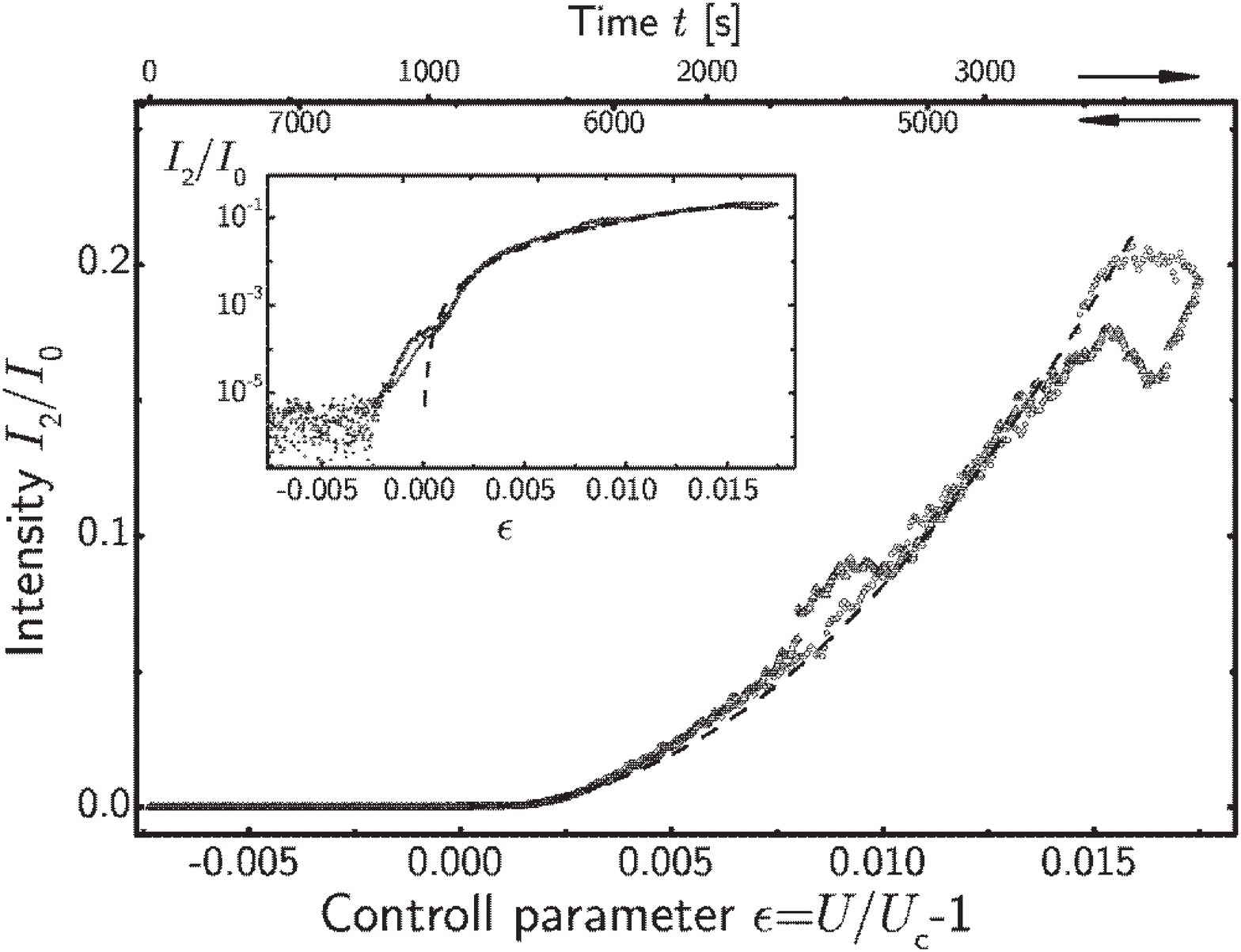}
\end{center}
\caption{ Intensity at the second order peak within a sequence of increasing
($\circ$) and decreasing ($\triangle$) driving voltage in cell 2. The fit to
a function $I \propto(U-U_{\text{c}})^{2}$ yields a critical voltage $U_{%
\text{c}}=28.84$\thinspace V at sine wave excitation with $\protect\nu= 500$%
\thinspace Hz. Due to a slow drift in conductivity the critical voltage is
shifted to a higher value as in Fig.\thinspace\protect\ref%
{figlaserdiffractionsim}. In the insert, the same data are presented on
logarithmic scale, an increase of fluctuation amplitudes in the subcritical
voltage range is clearly observable.}
\label{figbifurcation}
\end{figure}

For measurements near the instability threshold, we use cell 2. We study the
stationary director deflection amplitude as a function of the control
parameter $\epsilon=U/U_{\text{c}}-1$ by increasing the driving voltage
gradually from a subcritical value to a voltage above the threshold. The
voltage is increased adiabatically slowly such that the director deflection
is always practically in equilibrium. Simultaneously, the diffraction
intensity at the second order reflex is recorded as a quantitative measure
of the director field modulation. The constant scattering background is
eliminated as above. The experiment is repeated with the opposite direction
of the field sweep, the diffraction intensity is recorded while the driving
voltage is decreased with the same rate. The expected characteristics for a
perfect forward pitchfork bifurcation is $\varphi\propto\sqrt{\epsilon}$.
Considering relation (\ref{IvsPhi}), this would correspond to a quadratic
dependence of the diffraction intensity at the second order reflex from the
control parameter
\begin{equation}  \label{quad}
I_{2} \propto\varphi^{4} \propto\epsilon^{2} .
\end{equation}
Figure\thinspace\ref{figbifurcation} shows the experimental results. For
better visualization, data are presented in linear scale and logarithmically
in the insert. The systematic deviations of data taken during up and down
sweeps of the field, resp., are negligibly small. No hysteresis is found.
The parameter $U_{\text{c}}$ is fitted such to obtain best agreement with
Eq. (\ref{quad}). The measured intensities match the expected quadratic
behaviour, and therefore we associate the fitted $U_{\text{c}}$ with the
threshold voltage. Close to the threshold and below $U_{\text{c}}$, the
characteristics is covered by additional influences of noise. It clearly
deviates from the prediction of Eq. (\ref{quad}) indicated by the dashed
curve. One of the possible reasons is that subcritical fluctuations of the
modes close to the instability threshold \cite{Bisang98,Bisang99,Rehberg91}
lead to an increased diffraction signal at the corresponding position. In
addition, the equations used in the hydrodynamic model use exact planar
boundary conditions, while the cell actually has a small pretilt, typical
for glass plates of sandwich cells with antiparallel rubbing.

\subsection{Measurement of growth and decay rates at periodic excitation}

The solution of the linearized differential equations (\ref{growthrate1})
yield an exponential growth or decay of $\varphi_{t}$ (\ref{lyapunov1}) in
an electric field of constant amplitude. Equation (\ref{IvsPhi}) connects
this amplitude of the director deflection with the measurable diffraction
intensity, e.g. at the second order reflex. Both equations can be combined
to
\begin{equation}
I_{2}(t)=I_{2}(0)\text{e}^{4\nu\lambda_{1}\,t}=I_{2}(0)\text{e}^{\lambda _{%
\text{exp}}\,t},   \label{IvsLambda}
\end{equation}
where $\lambda_{1}(E,\nu,k_{\text{c}})$ is the largest eigenvalue of the
matrix product $\mathbf{T}^{+}\mathbf{T}^{-}$ and $\lambda_{\text{exp}}$ the
experimentally determined growth rate from the intensity change at the
second order reflex. The factor 4 considers the fourth order dependence of
the scattering intensity from the director deflection amplitude. Positive
growth rates for $\lambda_{\text{exp}}>0$ can be obtained in the experiment
by recording the intensity change at the reflex after an electrical field $%
E>E_{\text{c}}$ is turned on. For the measurement of $\lambda_{\text{exp}}<0$%
, the electric field is first switched to a supercritical value $E>E_{\text{c%
}}$ where the convection pattern develops. Then, the field is suddenly
changed to a value $E<E_{\text{c}}$ and the intensity trace is recorded.
Examples for these procedures in cell 1 are shown in Fig.\thinspace \ref%
{figgrowdecaysamples}. In order to detect the fast changes we use a
analog-digital-converter with a lower resolution (12\thinspace bit) but much
higher sampling frequency of 1\thinspace kHz. The constant background is
eliminated and the data are fitted to exponential functions in the middle of
the detection range. Growth and decay rates of 200 measurements are depicted
in Fig.~\ref{figgrowdecayrates} together with the eigenvalue $\lambda_{1}(E)$
calculated analytically from the material parameter in Table 1, and the
factor 4 from Eq. (\ref{IvsLambda}), has been taken into account. There is
good quantitative agreement with the linearized theory for electroconvection
for driving voltages around the threshold.

\begin{figure}[ptb]
\begin{center}
\includegraphics[width=8.5cm]{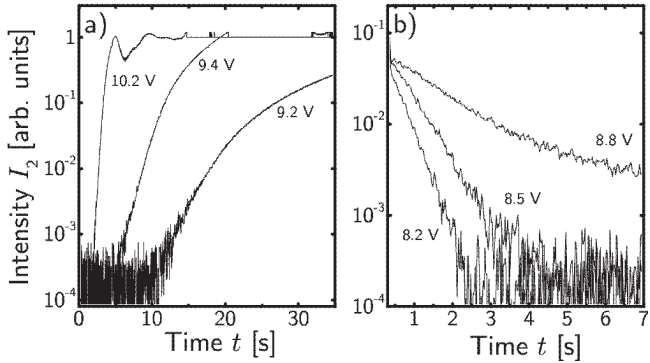}
\end{center}
\caption{ Time evolution of the diffraction intensity at the second order
reflex after changes of the field (20\thinspace Hz). The vertical axis shows
the amplified voltage at the photo diode, which is proportional to the
intensity $I_{2}$. The threshold $U_{\text{c}}$ is 9~V in case of cell 1. In
(a), the applied square wave voltage is switched at $t=0$ from zero to
supercritical voltages. In (b), the excitation voltage is switched down to
slightly subcritical values.}
\label{figgrowdecaysamples}
\end{figure}

\begin{figure}[ptb]
\begin{center}
\includegraphics[width=8.5cm]{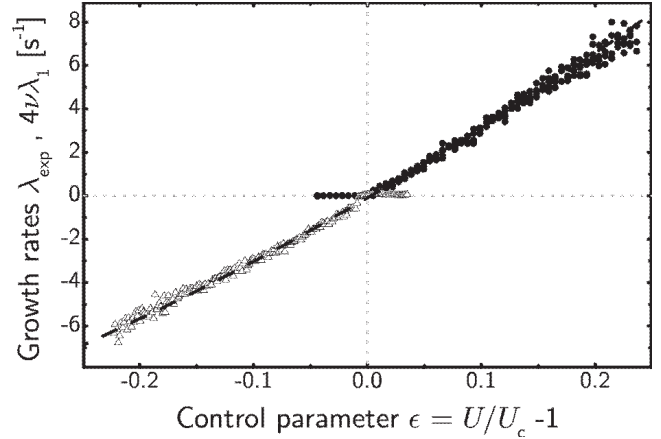}
\end{center}
\caption{ Growth and decay rates at different excitation amplitudes (square
wave). Growth rates ($\bullet$) has been obtained from traces similar to
those in Fig. \protect\ref{figgrowdecaysamples}a, and decay rates ($\triangle
$) from traces like those in Fig.\thinspace\protect\ref{figgrowdecaysamples}%
b. The dashed line depicts the theoretical prediction $4\protect\nu\protect%
\lambda_{1}(E)$ from a calculation of the largest eigenvalue of $\mathbf{T}%
^{+}\mathbf{T}^{-}$ in Eq. (\protect\ref{qpsisquare}). Material parameters
are given in Table 1.}
\label{figgrowdecayrates}
\end{figure}

\subsection{Trajectories under stochastic excitation}

\begin{figure}[ptb]
\begin{center}
\includegraphics[width=8.5cm]{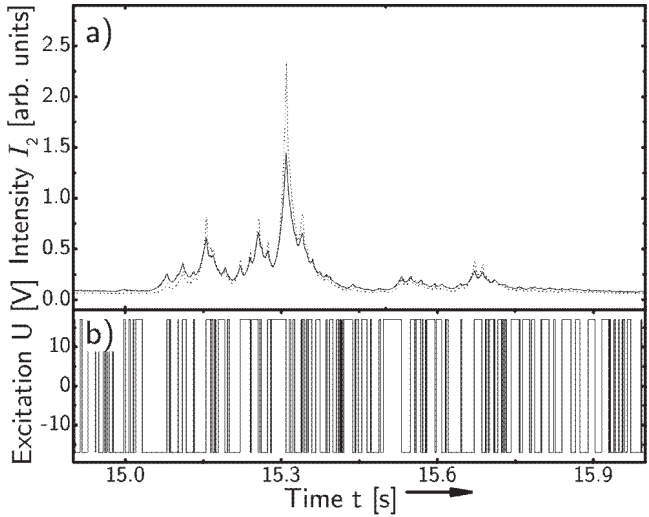}
\end{center}
\caption{ Real time detection of director amplitudes in stochastically
driven EHC, compared with a numerically simulated trajectory with the same
realization of the driving process. The upper part (a) shows a measured
intensity detected with the photodiode at the second order diffraction
reflex (solid line), and a simulated curve (dotted line), both corresponding
to the driving sequence depicted in (b). }
\label{figstochrates}
\end{figure}

If the deterministic voltage is replaced by a stochastic excitation
sequence, the trajectory of the director deflection exhibits irregular
changes in time. A statistical analysis of this phenomenon has been
described in detail \cite{johnPRE,johnPRL}. The measurement of the time
dependent diffraction intensity provides a convenient tool to study the
trajectories of pattern amplitudes in real time. Figure \ref{figstochrates}
shows the example of measured intensities at the second order reflex and the
corresponding numerical simulation of the trajectory by solution of the
differential equation (\ref{growthrate1}).

The stochastic sequence in this experiment was a dichotomous Markov process
(DMP) with jump rate 160\thinspace s$^{-1}$. Since the excitation sequence
is synthesized with a computer, it is possible to use identical noise
sequences in both experiment and simulation. In the bottom part of Fig.~\ref%
{figstochrates}, the realization of the stochastic driving process is shown.
The numerical $I(t)$ have been obtained from the $\varphi(t)$ trajectories
by use of (\ref{IvsPhi}). Figure~\ref{figstochrates} demonstrates that
experiment and theory for stochastically excited EHC do not only agree on
the statistical level when fundamental scaling laws are compared, but even
in details of the trajectories to a satisfactory degree, when we take into
account that the simulation cannot treat the involved additive noise exactly
but substitutes it by some average \cite{johnPRE}. Of course, even repeated
measurements of experimental trajectories are not exactly reproducible
because of such additive (thermal) noise. Trajectories taken with the same
noise sequence of the driving field differ in detail at small intensities,
but above the noise level they are very similar and reproduce the simulated
curve on average.

\section{Summary}

We have used laser diffraction as an detection method for direct
quantitative determination of the amplitudes of the director field in
nematic electroconvection. Instead of the evaluation of the complete
diffraction pattern, it is sufficient to record the second order diffraction
spot, which is mainly influenced by the phase grating generated by the
director field. An analytic treatment of the ray propagation in the LC layer
by Fermat's principle provides the qualitative relation $I_{2}/I_{0}=\text{%
const.}\times\varphi_{t}^{4}$ at the second order diffraction reflex for
small pattern amplitudes.

The proportionality constant can be derived from the material parameters
using an analytical approach that considers at least the first non-linear
term in light deflection in the calculation of the optical path of
individual light rays passing the cell. A numerical calculation of light
propagation, which does not use mathematical approximations, except for the
concept of ray optics, confirms the analytical result up to sufficiently
large director deflections. The complete diffraction intensity profile
calculated numerically is in good agreement with the profile measured
experimentally. It has been shown that it is sufficient to use the simple
relation Eq. (\ref{IvsPhi}) to determine the absolute value of the director
deflection amplitude from the diffraction efficiency. We note, however, that
a rigorous treatment of the problem of light propagation (for example by
means of the FTDT method \cite{Taflove}) is the only exact treatment of the
optical problem. It has not been achieved yet.

The derived quantitative relations have been used in three applications. In
case of periodic sine wave excitation, the reported technique permits us to
confirm the square root characteristics of the pattern amplitude in the
pitchfork bifurcation of the stripe pattern. The non perfect behaviour at
subcritical values is attributed to thermal fluctuations and a slight sample
pretilt.

The agreement between analytically calculated growth and decay rates of the
amplitudes of the director field and the measured light intensity at the
second order reflex of diffracted light shows that one has to be very
cautious when growth/decay rates are determined from laser diffraction
intensities. The decay of the optical signal goes with the 4th power of the
director deflections, and consequently, time constants differ by a factor of
four.

The real-time quasi optical \textsc{Fourier} transformation of the pattern
gives an easy access to fast changing mode spectra or amplitudes, e.g. in
case of stochastic driving. We have demonstrated the direct correlation
between the driving electric field and the response of the director.
Observation of the whole diffraction image instead of the trajectory of one
representative diffraction peak may provide access to the development of the
mode spectrum and access to dynamic mode selection in the stochastic driven
system.

An important aspect in the experiment is the strong dependence of the
diffraction image on small deviations from normal incidence. The problem of
oblique incidence has been addressed first by Zenginoglou and Kosmopoulos
\cite{Zengi88}. In the appendix, we consider in detailed the dependence of
the diffraction pattern from the angle of incidence of the laser beam. It is
demonstrated that oblique incidence in general favors the reflexes of odd
numbered order, which in first line reflect the amplitude grating produced
by the director field.

The authors are particularly indebted to H. Schmiedel for helpful
comments, discussions and critical reading of the manuscript. We acknowledge
financial support from the Deutsche Forschungsgemeinschaft (Grant Be 1417/4 and
SFB 294).

\begin{table}[ptb]
\begin{center}
\begin{tabular}{llr@{.}lr@{.}l}
\hline\hline
\multicolumn{2}{c}{Parameter} & \multicolumn{2}{c}{cell 1 (2)} &
\multicolumn{2}{c}{Exper. value} \\ \hline
$n_{\text{o}}$ &  & 1 & 4935 & 1 & 4935 \\
$n_{\text{e}}$ &  & 1 & 6315 & 1 & 6315 \\
$\varepsilon_{\scriptscriptstyle\parallel}$ &  & 6 & 24 & 6 & 24 \\
$\varepsilon_{\scriptscriptstyle\perp}$ &  & 6 & 67 & 6 & 67 \\
$\sigma_{\scriptscriptstyle\parallel}$ & $[\mbox{s}^{-1}]$ & 90 & 0 (1350) &
117 & 0 \\
$\sigma_{\scriptscriptstyle\perp}$ & $[\mbox{s}^{-1}]$ & 60 & 0 (900) & 90 &
0 \\
$\alpha_{1}$ & $[\mbox{g\thinspace cm}^{-1}\mbox{s}^{-1}]$ & 0 & 1 &
\multicolumn{2}{l}{} \\
$\gamma_{1}$ & $[\mbox{g\thinspace cm}^{-1}\mbox{s}^{-1}]$ & 3 & 3 & 3 & 6
\\
$\gamma_{2}$ & $[\mbox{g\thinspace cm}^{-1}\mbox{s}^{-1}]$ & -3 & 3 &
\multicolumn{2}{l}{} \\
$\beta$ & $[\mbox{g\thinspace cm}^{-1}\mbox{s}^{-1}]$ & \multicolumn{2}{l}{}
& \multicolumn{2}{l}{} \\
$\eta_{1}$ & $[\mbox{g\thinspace cm}^{-1}\mbox{s}^{-1}]$ & 3 & 62 &
\multicolumn{2}{l}{} \\
$\eta_{2}$ & $[\mbox{g\thinspace cm}^{-1}\mbox{s}^{-1}]$ & 1 & 0 &
\multicolumn{2}{l}{} \\
$K_{11}$ & $[\mbox{g\thinspace cm\thinspace s}^{-2}]$ & 14 & 9 $%
\times10^{-7} $ & 14 & 9$\times10^{-7}$ \\
$K_{33}$ & $[\mbox{g\thinspace cm\thinspace s}^{-2}]$ & 13 & 76 $\times
10^{-7}$ & 13 & 76$\times10^{-7}$ \\ \hline\hline
\end{tabular}%
\end{center}
\caption{Material parameters in Eq. (\protect\ref{qphionestep}) used in the
calculations. Experimentally data for \emph{Mischung 5} (last column) have
been taken from \protect\cite{Amm99,MGRS}, measured conductivities
correspond to the non-doped material. The unknown parameters and
conductivities of the individual cells are obtained by fitting the
corresponding threshold voltages and wave number characteristics for
periodic ac driving to experimental data, see also Fig.~{\protect\ref%
{figthresholds}}. }
\label{tabmaterial}
\end{table}

\bibliographystyle{prsty}
\bibliography{laser}

\begin{appendix}
\section{Diffraction at oblique incidence}
\begin{figure}[ptb]
\begin{center}
\includegraphics[width=8.5cm]{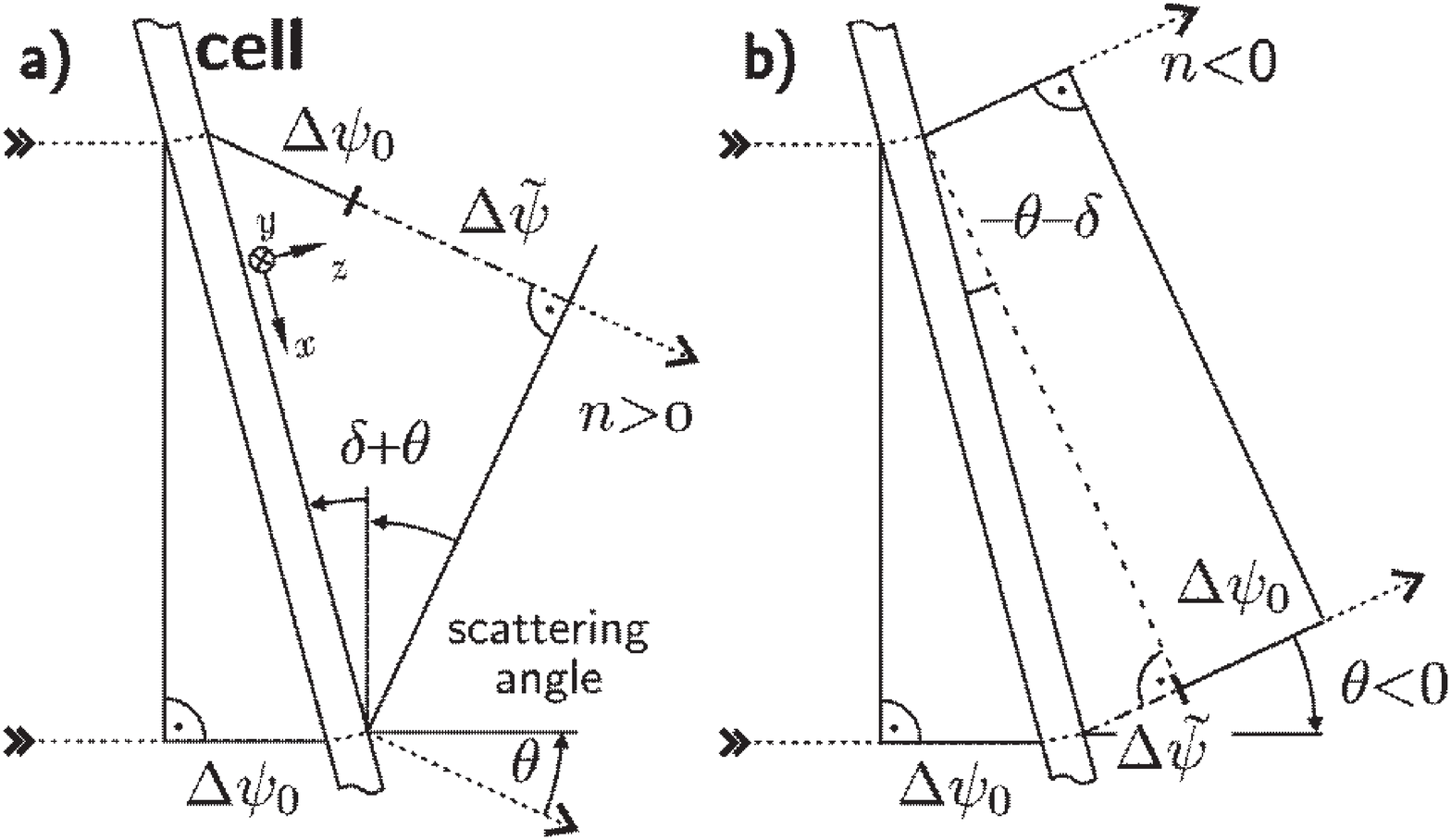}
\end{center}
\caption{ A sketch of two light rays under oblique incidence seperated by the wave length of the phase grating wave
length $\lambda_{\text{ph}}$. The additional phase $\Delta\psi^{\prime}$ has to be considered in the calculation of the
\textsc{Frauenhofer} diffraction for the positive (a) and negative (b) wings
of the diffraction image. }%
\label{Figrotcellsketch}%
\end{figure}

\begin{figure}[ptb]
\begin{center}
\includegraphics[width=8.5cm]{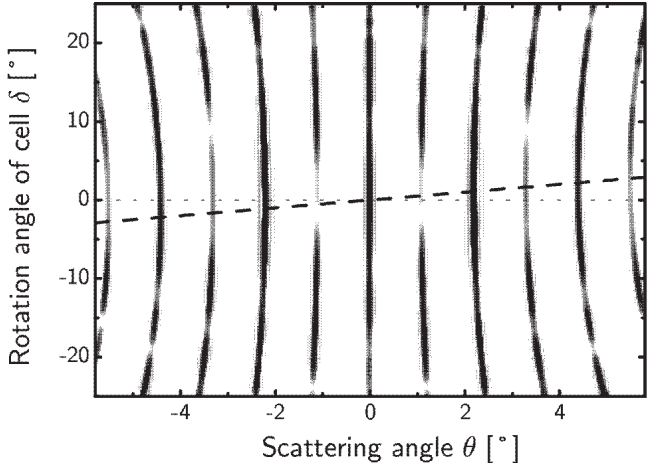}
\end{center}
\caption{Density plot from simulated diffraction profiles for oblique incidence and a $\varphi=0.4$~rad. The gray scale
is logarithmic in the intensities. A cell thickness of 25\thinspace
\textmu m and a pattern wave length of 33\thinspace \textmu m assumed. The
dashed line marks $\delta =\frac{1}{2}\theta$.
}%
\label{figrotcellsim}%
\end{figure}

The experiments show that the diffraction profile has a strong dependence
on a the tilt of the cell respective to the incident laser beam. A
theoretical treatment with linearization has been published in
\cite{Zengi88}. In case of oblique incidence, where the cell is tilted in
the ($x,z$) plane, the symmetry $\theta\leftrightarrow-\theta$ is broken.
The initial condition for Eq. (\ref{fullEulerDGL}) is now
$r^{\prime}(z)=\tan\delta,$ where $\delta$ is the entry angle of the laser
beam into the LC-layer. We consider this in the numerical calculation of
the beam propagation. In addition to this symmetry breaking,  an extra
phase $\Delta\tilde{\psi}(\theta)$ appears. Figure~\ref{Figrotcellsketch}
illustrates the origin of the additional phase difference. For the primary
beam at $\theta=0$, $\Delta\tilde{\psi}(\theta)$ vanishes. The symmetry
breaking leads to a slight shift of the diffraction spots, and more
obviously, to a change of the relative intensities of even and odd order
spots. The condition allowing for the additional phase for constructive
interference on the 2n$th$ even order spot ($n>0,
\theta_n>0$) is (see Fig. \ref{Figrotcellsketch}a)
\begin{gather}
\lambda_{\text{ph}}\sin(\theta_{n}+\delta)=\Delta\psi_{0}
+n\frac{2\pi}{k_{\text{\tiny L}}},\\%
\theta_{n}=\arcsin\left( \frac{2\pi n}{\lambda_{\text{ph}} k_{\text{\tiny L}}}
+\sin\delta\right)  -\delta\;;\;n>0
\end{gather}
and for $\theta_n<0$%
\begin{equation}
\theta_{n}=\arcsin\left(  \frac{2\pi |n|}{\lambda_{\text{ph}} k_{\text{\tiny L}}}
-\sin\delta\right)+\delta\;;\;n<0,
\end{equation}
where $\lambda_{\text{ph}}$ is the wave length of the phase modulation of exiting light and $k_{\text{\tiny L}}$ the
wave number of the laser light.
The diffraction intensity calculated numerically for a director modulation of $\varphi=0.4$~rad as a function of
$\theta$ and $\delta$ is depicted in the density plot of Fig. \ref{figrotcellsim}. The most obvious result of the
numerical calculation are the quantitative changes of the diffraction intensities with the cell rotation angle. Whereas
the odd order maxima, which are mainly generated by the amplitude modulation, exhibit a minimum in the non-tilted cell
and increased with a slight tilt of the cell, even order spots show qualitatively opposite behaviour. In the experiment
we record the complete diffraction image on a diffusely scattering screen with a CCD camera (see Fig.\thinspace
\ref{figsetup}) and scan the line $y=0$ (see Fig.\ref{figsampleimages}a) from the digital image sequence. The camera
gives only a qualitative picture of the intensities, not an exactly linear representation, but qualitative agreement
with Fig.~\ref{figrotcellsim} is clearly acknowledged. We remark that the smallest diffraction angle $\theta_n(\delta)$
for a given order $n$ is not reached at $\delta=0$, but
\begin{equation}
\frac{\text{d}}{\text{d}\delta}\theta_{n}=0\;\Rightarrow\;\sin\delta =\frac{1}{2}\frac{2\pi n}{\lambda_{\text{ph}}
k_{\text{\tiny L}}} \approx \frac{1}{2}\sin\theta.
\end{equation}
Its position is depicted as dashed line in Fig.~\ref{figrotcellsim} and coincides with the largest
amplification/attenuation, resp., of the diffraction peak intensities.

\begin{figure}[ptb]
\begin{center}
\includegraphics[width=8.5cm]{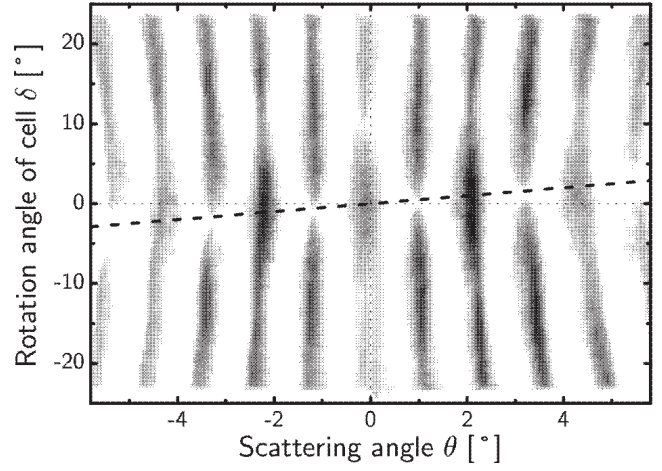}
\end{center}
\caption{Density plot of the measured diffraction profiles in inverse gray scale for a tilted cell. The profiles are
taken in the midplane $y=0$ of the diffraction images of a stable convection pattern. The tilt of the cell to the
incident beam leads to modulations of relative intensities and positions for all reflexes. One acknowledges the
symmetry $(-\delta,-\theta)\leftrightarrow(\delta,\theta)$.
The dashed line marks $\delta =\frac{1}{2}\theta$ as in Fig. \ref{figrotcellsim}.}%
\label{figrotcellexp}%
\end{figure}
\end{appendix}

\end{document}